\documentclass
[aps,pra,amsfonts,amssymb,twocolumn,amsmath,preprintnumbers,nofootinbib,floatfix,
showpacs,superscriptaddress]{revtex4-1}%
\usepackage[dvips]{graphics}
\usepackage{graphicx}
\usepackage{dcolumn}
\usepackage{bm}
\usepackage{amsmath}
\usepackage{amsfonts}
\usepackage{amssymb}
\usepackage{xcolor}
\usepackage{subfigure}
\usepackage{hyperref,hypcap}
\usepackage{braket}
\usepackage{commath}%
\providecommand{\U}[1]{\protect\rule{.1in}{.1in}}
\begin{document}

\title{Theory of nonlinear Hall effects: renewed semiclassics from quantum kinetics}
\author{Cong Xiao}
\affiliation{Department of Physics, The University of Texas at Austin, Austin, Texas 78712, USA}

\author{Z. Z. Du}
\affiliation{Shenzhen Institute for Quantum Science and Engineering and Department of Physics,
Southern University of Science and Technology, Shenzhen 518055, China}
\affiliation{Shenzhen Key Laboratory of Quantum Science and Engineering, Shenzhen 518055, China}
\affiliation{Peng Cheng Laboratory, Shenzhen 518055, China}

\author{Qian Niu}
\affiliation{Department of Physics, The University of Texas at Austin, Austin, Texas 78712, USA}

\begin{abstract}
We propose a modified Boltzmann nonlinear electric-transport framework which
differs from the nonlinear generalization of the linear Boltzmann formalism by
a contribution that has no counterpart in linear response. This contribution
follows from the interband-coherence effect of dc electric-fields during
scattering and is related to the interband Berry connection. As an
application, we demonstrate it in the second-order nonlinear Hall effect of
the tilted massive Dirac model. The intuitive Boltzmann constructions are
confirmed by a quantum kinetic theory, which shows that arbitrary $n$th-order
nonlinear dc response up to the first three leading contributions in the weak
disorder potential is handled by the same few gauge-invariant semiclassical ingredients.

\end{abstract}
\maketitle

%\author{Cong Xiao}
%\email{congxiao@utexas.edu}
%\affiliation{Department of Physics, The University of Texas at Austin, Austin, Texas 78712, USA}

%\author{Z. Z. Du}
%\affiliation{Shenzhen Institute for Quantum Science and Engineering and Department of Physics,
%Southern University of Science and Technology, Shenzhen 518055, China}
%\affiliation{Shenzhen Key Laboratory of Quantum Science and Engineering, Shenzhen 518055, China}
%\affiliation{Peng Cheng Laboratory, Shenzhen 518055, China}

%\author{Qian Niu}
%\affiliation{Department of Physics, The University of Texas at Austin, Austin, Texas 78712, USA}

\section{Introduction}

The nonlinear response to an applied electric field in crystalline solids has
attracted revived interest, owing to the essential role played by the quantum
geometry of the Bloch wave function
\cite{Nagaosa2018,Moore2010,Gao2014,Fu2015}. In the optical high-frequency
regime of the electric field, the shift current photogalvanic effect and
second Harmonic generation have been shown to be related to the Berry
connection of each involved Bloch band \cite{Morimoto2016,Wu2017}. In the
low-frequency regime, higher-order moments of the Berry curvature in momentum
space emerge in the nonlinear anomalous Hall responses in the absence of
magnetic field, such as the Berry curvature dipole
\cite{Fu2015,Xu2018,Facio2018,Lu2018,Zhang2018,Yan2018,Low2018,Low2015,Zhou2019,Souza2018}
and quadrupole \cite{Parker2019} in second- and third-order Hall responses,
respectively. In particular, the second-order nonlinear response dominates the
anomalous Hall effect in time-reversal-invariant crystals that break inversion
symmetry, and has been observed in few-layer WTe$_{2}$ \cite{Ma2019,Mak2019}.

The quantum geometry of the Bloch electron also influences its scattering with
disorder. A prominent case is the linear anomalous Hall effect
\cite{Nagaosa2010}, where nonzero Berry connection and curvature imply the
presence of two asymmetric scattering effects termed as skew scattering and
side-jump \cite{Sinitsyn2006}. A Boltzmann transport formalism for the linear
anomalous Hall effect has been established
\cite{Sinitsyn2008,Sinitsyn2007,Luttinger1958,Ado2015,Xiao2018KL}. It has been
generalized phenomenologically in the recent efforts to understand the
second-order nonlinear response in the low-frequency limit
\cite{Deyo2009,Konig2019,Fu2018,Du2018}. A basic question then arises: Is this
framework valid in nonlinear responses? The existing quantum transport
theories \cite{Sodemann2019,Morimoto2016PRB} have not settled this issue.
Moreover, there exist two different proposals \cite{Deyo2009,Fu2018} to
generalize phenomenologically the side-jump contribution to the second-order
nonlinear response. More importantly, it should be worried whether there is
other contribution that is missed in the direct generalization of the
aforementioned semiclassical formalism. If there is, can the nonlinear
response still be grasped by the few gauge-invariant semiclassical ingredients
as in the linear response?

In this work we address all the above concerns in the dc limit by developing a
recursive quantum kinetic theory for arbitrary $n$th-order (finite $n$)
electric current response. We focus on the first three leading order
contributions in the weak disorder potential $\hat{V}$ (namely, $V^{-2n}$,
$V^{-2n+1}$, $V^{-2n+2}$ in the $n$th-order electric transport), which are
usually sufficient to account for both the longitudinal and transverse
transport in the regime $\hbar/\tau<\Delta$ ($\tau$ is the scattering time,
$\Delta$ is the band splitting around the Fermi level). Remarkably, we find
that arbitrary order nonlinear response retains the same structure as the
linear response, except for a contribution resulting from the electric-field
induced interband virtual transition during the scattering. This contribution
only contributes to nonlinear response and is related to the interband Berry
connection. A modified Boltzmann nonlinear-response framework thus emerges,
establishing for the first time the consistency between the renewed Boltzmann
and quantum kinetics in nonlinear (Hall) electric transport. As an
application, we show the aforementioned contribution in the second-order
nonlinear Hall effect of the two-dimensional (2D) tilted massive Dirac model.

Our paper is organized as follows. In Sec. II we set forth the renewed
Boltzmann theory for nonlinear electric transport, which is applied to the
model calculation of the second-order nonlinear Hall effect in Sec. III. The
quantum kinetic theory that underlies the Boltzmann formulation is outlined in
Sec. IV, with the main ideas and results elaborated. Finally, we compare our
theory to other existing theories in Sec. V and conclude this paper in Sec.
VI. The detailed derivation of our quantum kinetic theory is presented in the
Supplemental Material \cite{Supp} for the convenience of interested readers.

\section{Renewed semiclassics}

The outcomes of the quantum kinetic approach are found to correspond to a
semiclassical Boltzmann way to understand the nonlinear transport. In this
section we first describe the latter framework, considering its great physical
transparency and simplicity.

In the Boltzmann description of electronic transport in crystalline solids,
the charge current density is given by%
\begin{equation}
\bm j=e\sum_{l}F_{l}\bm v_{l},
\end{equation}
where the occupation function $F_{l}$ of the Bloch-state $|l\rangle=|\eta\bm
k\rangle$, with $\eta$ the band index and $\bm k$ the crystal momentum, and
the velocity $\bm v_{l}$ are two central quantities. In the perturbative
treatment for the weak electric field, $F_{l}$ can be expanded in terms of
ascending powers (denoted by $n$) of the electric field $\bm E$, namely%
\begin{equation}
F_{l}=\sum_{n\geq0}F_{n,l},\label{Fn}%
\end{equation}
where $F_{n,l}\propto E^{n}$\ is the occupation function responsible for the
$n$th-order electric transport.

In the conventional Boltzmann recipe the driving term by the applied electric
field and the collision term by scattering are clearly separated in the
steady-state Boltzmann equation \cite{Ziman1972}
\begin{equation}
-\frac{e}{\hbar}\bm E\cdot\partial_{\bm k}F_{l}=\sum_{l^{\prime}}%
(\omega_{l^{\prime}l}^{\left(  2\right)  }F_{l}-\omega_{ll^{\prime}}^{\left(
2\right)  }F_{l^{\prime}}).\label{SBE}%
\end{equation}
The semiclassical scattering rate, regarded to be independent of the electric
field, is given by the golden rule
\begin{equation}
\omega_{l^{\prime}l}^{\left(  2\right)  }=\omega_{ll^{\prime}}^{\left(
2\right)  }=\frac{2\pi}{\hbar}W_{l^{\prime}l}\delta\left(  \epsilon
_{l}-\epsilon_{l^{\prime}}\right)  ,\text{ \ }W_{l^{\prime}l}=\langle
\left\vert V_{ll^{\prime}}\right\vert ^{2}\rangle_{c}.\label{Born}%
\end{equation}
Here $W_{l^{\prime}l}$ is the scattering matrix element, $\left\langle
..\right\rangle _{c}$ stands for the disorder average. In the constant
relaxation time approximation the collision term on the right hand side of Eq.
(\ref{SBE}) reduces to $F_{l}/\tau$, where $1/\tau\sim V^{2}$, and the
recursive solution of this equation yields the scaling $F_{n,l}\sim E^{n}%
\tau^{n}\sim E^{n}V^{-2n}$.

\subsection{Modification to scattering by electric field}

In the conventional Boltzmann equation the scattering process is independent
of the electric field. But this is not true in general. A prominent example is
the linear anomalous Hall current originating from the work done by the
electric field during scattering. The key ingredient here is the
coordinate-shift of a semiclassical electron during any scattering process
\cite{Sinitsyn2006}%
\begin{equation}
\delta\bm{r}_{l^{\prime}l}=\mathcal{A}_{l^{\prime}}-\mathcal{A}_{l}-\left(
\partial_{\bm{k}}+\partial_{\bm{k}^{\prime}}\right)  \arg V_{l^{\prime}l},
\label{shift}%
\end{equation}
where $\mathcal{A}_{l}=\langle u_{\eta\bm k}|i\partial_{\bm k}|u_{\eta\bm
k}\rangle$ is the intraband Berry connection, with $|u_{\eta\bm k}\rangle$ the
periodic part of the Bloch state. This picture implies that the energy
conservation condition in the golden rule [Eq. (\ref{Born})] is modified to
be
\begin{equation}
\delta\left(  \epsilon_{l}-\epsilon_{l^{\prime}}+e\bm E\cdot\delta
\bm r_{l^{\prime}l}\right)  \simeq\delta\left(  \epsilon_{l}-\epsilon
_{l^{\prime}}\right)  +\frac{\partial\delta\left(  \epsilon_{l}-\epsilon
_{l^{\prime}}\right)  }{\partial\epsilon_{l}}e\bm E\cdot\delta\bm r_{l^{\prime
}l}. \label{delta}%
\end{equation}
The direct generalization of this semiclassical construction into nonlinear
responses leads to an occupation function which scales as $F_{n,l}\sim
E^{n}\tau^{n-1}\sim E^{n}V^{-2n+2}$. This $F_{n,l}$ yields an important
contribution to the second-order nonlinear Hall effect
\cite{Du2018,Konig2019,Deyo2009}. Note that the first-order expansion in the
above equation is already sufficient to obtain the $F_{n,l}$ of order of
$V^{-2n+2}$.

We reveal in the following that, there is another electric-field-induced
effect during scattering, which only contributes to nonlinear responses. The
intuitive motivation is that not only the energy conservation delta-function
but also the scattering matrix element $W_{l^{\prime}l}$ of the semiclassical
scattering rate should be corrected by the $\bm E$-field. This term is an
interband-coherence (interband virtual transition) effect of the $\bm E$-field
during scattering. More precisely, the Bloch states involved in the scattering
have to be dressed by the electric field, thus $V_{ll^{\prime}}\rightarrow
\langle\tilde{l}|\hat{V}|\tilde{l}^{\prime}\rangle$ where $|\tilde{l}%
\rangle=|l\rangle+|\delta^{\bm E}l\rangle$ is the $\bm E$-field-dressed Bloch
state, and%
\begin{equation}
|\delta^{\bm E}l\rangle=-e\sum_{l^{\prime\prime}}^{\prime}|l^{\prime\prime
}\rangle\frac{\bm E\cdot\mathcal{A}_{l^{\prime\prime}l}}{\epsilon_{l}%
-\epsilon_{l^{\prime\prime}}}\label{mixture}%
\end{equation}
arises from the electric-field induced interband virtual transition
\cite{Xiao2017SOT-SBE}. $\mathcal{A}_{ll^{\prime}}=\langle u_{\eta\bm
k}|i\partial_{\bm k}|u_{\eta^{\prime}\bm k}\rangle$ is the interband Berry
connection. Hereafter the notation $\sum^{\prime}$ means that all the index
equalities should be avoided in the summation. In order to obtain the
$F_{n,l}$ of order of $V^{-2n+2}$, it is sufficient to retain
\begin{equation}
W_{l^{\prime}l}\rightarrow W_{l^{\prime}l}+\delta^{\bm E}W_{l^{\prime}%
l},\label{W}%
\end{equation}
where the $\bm E$-field corrected scattering matrix element is linear in $\bm
E$\ and reads
\begin{align}
\delta^{\bm E}W_{l^{\prime}l} &  =2\operatorname{Re}\langle V_{ll^{\prime}%
}(\langle l^{\prime}|\hat{V}|\delta^{\bm E}l\rangle+\langle\delta^{\bm
E}l^{\prime}|\hat{V}|l\rangle)\rangle_{c}\label{new term}\\
&  =-e\bm E\cdot\sum_{l^{\prime\prime}}^{\prime}2\operatorname{Re}\langle
\frac{V_{ll^{\prime}}V_{l^{\prime}l^{\prime\prime}}\mathcal{A}_{l^{\prime
\prime}l}}{\epsilon_{l}-\epsilon_{l^{\prime\prime}}}+\frac{V_{ll^{\prime}%
}\mathcal{A}_{l^{\prime}l^{\prime\prime}}V_{l^{\prime\prime}l}}{\epsilon
_{l^{\prime}}-\epsilon_{l^{\prime\prime}}}\rangle_{c}.\nonumber
\end{align}

\begin{figure}[tbh]
\includegraphics[width=0.9\columnwidth]{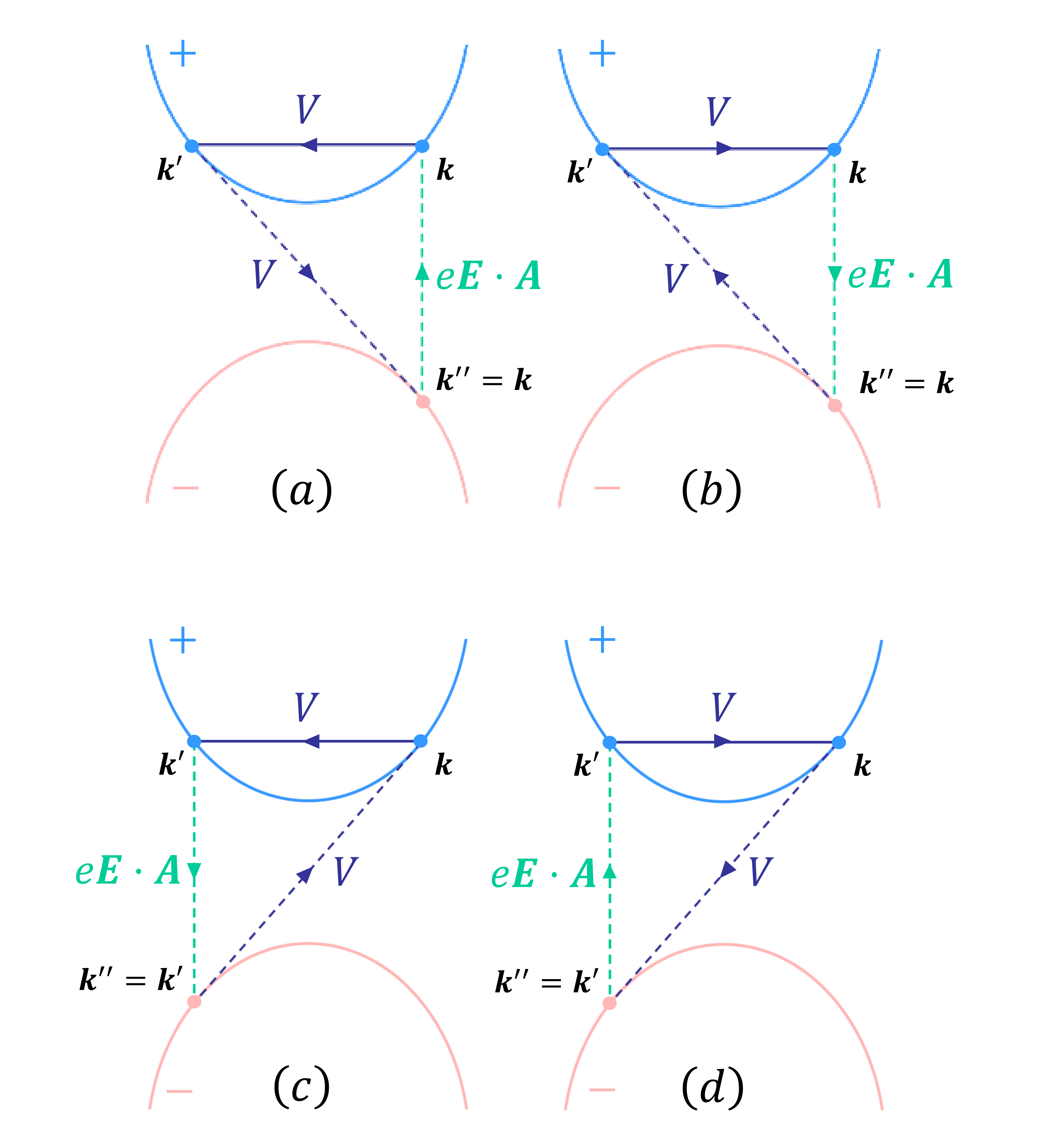}\caption{{}Schematics
of Eq. (\ref{new term}) describing the electric-field induced interband
virtual processes during scattering in two-band systems. The contributions
from (a) and (b) to $\delta^{\bm E}W_{l^{\prime}l}$ are complex conjugated, so
do (c) and (d). }%
\label{fig:new-term}%
\end{figure}

In Fig.~\ref{fig:new-term} we show schematically the physical processes
described by $\delta^{\bm E}W_{l^{\prime}l}$ in a two-band system. The Fermi
level is assumed to locate at the conduction band. Because of the presence of
the vertical interband virtual transition induced by the electric field, these
scattering processes involve an off-shell Bloch state away from the Fermi
surface. 

Collecting Eqs. (\ref{Born}), (\ref{delta}), (\ref{W}) and (\ref{new term}),
the $\bm E$-field corrected scattering rate takes the following form
\begin{equation}
\delta^{\bm E}\omega_{l^{\prime}l}^{\left(  2\right)  }=\delta_{1}^{\bm
E}\omega_{l^{\prime}l}^{\left(  2\right)  }+\delta_{2}^{\bm E}\omega
_{l^{\prime}l}^{\left(  2\right)  },\label{1+2}%
\end{equation}
where%
\begin{equation}
\delta_{1}^{\bm E}\omega_{l^{\prime}l}^{\left(  2\right)  }=\frac{2\pi}{\hbar
}W_{l^{\prime}l}\frac{\partial\delta\left(  \epsilon_{l}-\epsilon_{l^{\prime}%
}\right)  }{\partial\epsilon_{l}}e\bm E\cdot\delta\bm r_{l^{\prime}%
l},\label{1}%
\end{equation}%
\begin{equation}
\delta_{2}^{\bm E}\omega_{l^{\prime}l}^{\left(  2\right)  }=\frac{2\pi}{\hbar
}\delta^{\bm E}W_{l^{\prime}l}\delta\left(  \epsilon_{l}-\epsilon_{l^{\prime}%
}\right)  .\label{2}%
\end{equation}
While $\delta_{1}^{\bm E}\omega_{l^{\prime}l}^{\left(  2\right)  }$ has been
well-known, $\delta_{2}^{\bm E}\omega_{l^{\prime}l}^{\left(  2\right)  }$\ is
proposed for the first time in the context of the nonlinear Hall effect. We
note here that interband virtual processes are also indispensable in
$\delta_{1}^{\bm E}\omega_{l^{\prime}l}^{\left(  2\right)  }$ through the
coordinate-shift $\delta\bm r_{l^{\prime}l}$ \cite{Sinitsyn2006}.

\subsection{Boltzmann equation for nonlinear responses}

Taking into account the effect of the $\bm E$-field during scattering, the
Boltzmann equation (\ref{SBE}) is modified to be%
\begin{equation}
-\frac{e}{\hbar}\bm E\cdot\partial_{\bm k}F_{l}=\sum_{l^{\prime}}%
(\omega_{l^{\prime}l}^{\left(  2\right)  }+\delta^{\bm E}\omega_{l^{\prime}%
l}^{\left(  2\right)  })\left(  F_{l}-F_{l^{\prime}}\right)
.\label{E-corrected SBE}%
\end{equation}
In combination with Eq. (\ref{Fn}), it can be written in a recursive form%
\begin{gather}
-\frac{e}{\hbar}\bm E\cdot\partial_{\bm k}F_{n-1,l}-\sum_{l^{\prime}}%
\delta^{\bm E}\omega_{l^{\prime}l}^{\left(  2\right)  }\left(  F_{n-1,l}%
-F_{n-1,l^{\prime}}\right)  \nonumber\\
=\sum_{l^{\prime}}(\omega_{l^{\prime}l}^{\left(  2\right)  }F_{n,l}%
-\omega_{ll^{\prime}}^{\left(  2\right)  }F_{n,l^{\prime}}),\text{
\ \ \ \ \ \ }\left(  n\geq1\right)  \label{recursive SBE}%
\end{gather}
where the effect of the electric-field during scattering appears as an
effective driving term in the Boltzmann equation for $F_{n,l}$, and $F_{n,l}%
$\ is accurate to the third leading order of the weak disorder potential,
namely the order of $V^{-2n+2}$.

In linear response, $n=1$, $F_{0,l}$ is the Fermi distribution and
$(F_{0,l}-F_{0,l^{\prime}})\delta(\epsilon_{l}-\epsilon_{l^{\prime}})=0$, thus
$\delta_{2}^{\bm E}\omega_{l^{\prime}l}^{\left(  2\right)  }$ does not
contribute to the Boltzmann equation. This explains why this term is absent in
the Boltzmann theory of linear response \cite{Sinitsyn2008}. By contrast, it
constitutes a basic ingredient of the Boltzmann description of nonlinear responses.

Higher-order disorder corrections to $\omega_{l^{\prime}l}^{\left(  2\right)
}$ on the right hand side of Eq. (\ref{recursive SBE}) are included through
replacing $V_{ll^{\prime}}$ by the T-matrix $T_{ll^{\prime}}$
\cite{Sinitsyn2008,Xiao2018KL}. The golden rule thus yields $\omega
_{ll^{\prime}}=\omega_{ll^{\prime}}^{\left(  2\right)  }+\omega_{ll^{\prime}%
}^{\left(  3\right)  }+\omega_{ll^{\prime}}^{\left(  4\right)  }$ up to the
first three leading orders of the disorder potential. Hereafter the
superscript $\left(  i\right)  $\ means the order in disorder potential. It is
easy to check that such corrections to $\delta^{\bm E}\omega_{l^{\prime}%
l}^{\left(  2\right)  }$ are not needed provided that the electric current is
considered up to the three leading orders of the disorder potential.

In the considered case the occupation function is the sum of the leading (L),
sub-leading (SL) and sub-sub-leading (SSL) contributions:%
\begin{equation}
F_{n,l}=F_{n,l}^{\text{L}}+F_{n,l}^{\text{SL}}+F_{n,l}^{\text{SSL}},
\end{equation}
where $F_{n,l}^{\text{L}}$, $F_{n,l}^{\text{SL}}$ and $F_{n,l}^{\text{SSL}}$
are of order of $V^{-2n}$, $V^{-2n+1}$ and $V^{-2n+2}$, respectively. The
semiclassical occupation functions with positive exponent of $V$ can be
neglected in the weak disorder regime, thus in equilibrium $F_{0,l}%
=F_{0,l}^{\text{L}}$ is just the Fermi distribution and $F_{0,l}^{\text{SL}%
}=F_{0,l}^{\text{SSL}}=0$. This point is in fact implicit in previous works on
the semiclassical Boltzmann theories for the linear and nonlinear anomalous
Hall effects \cite{Deyo2009,Konig2019,Fu2018,Du2018,Sinitsyn2008}.

Therefore, the Boltzmann equation can be cast into the following three
equations%
\begin{equation}
-\frac{e}{\hbar}\bm E\cdot\partial_{\bm k}F_{n-1,l}^{\text{L}}=\sum
_{l^{\prime}}\omega_{l^{\prime}l}^{\left(  2\right)  }(F_{n,l}^{\text{L}%
}-F_{n,l^{\prime}}^{\text{L}}),\label{SBE-L}%
\end{equation}%
\begin{gather}
-\frac{e}{\hbar}\bm E\cdot\partial_{\bm k}F_{n-1,l}^{\text{SL}}=\sum
_{l^{\prime}}\omega_{l^{\prime}l}^{\left(  2\right)  }(F_{n,l}^{\text{SL}%
}-F_{n,l^{\prime}}^{\text{SL}})\nonumber\\
+\sum_{l^{\prime}}(\omega_{l^{\prime}l}^{\left(  3\right)  as}F_{n,l}%
^{\text{L}}-\omega_{ll^{\prime}}^{\left(  3\right)  as}F_{n,l^{\prime}%
}^{\text{L}}),\label{SBE-SL}%
\end{gather}
and%
\begin{gather}
-\frac{e}{\hbar}\bm E\cdot\partial_{\bm k}F_{n-1,l}^{\text{SSL}}%
-\sum_{l^{\prime}}\delta^{\bm E}\omega_{l^{\prime}l}^{\left(  2\right)
}(F_{n-1,l}^{\text{L}}-F_{n-1,l^{\prime}}^{\text{L}})=\nonumber\\
\sum_{l^{\prime}}\omega_{l^{\prime}l}^{\left(  2\right)  }(F_{n,l}%
^{\text{SSL}}-F_{n,l^{\prime}}^{\text{SSL}})+\sum_{l^{\prime}}(\omega
_{l^{\prime}l}^{\left(  4\right)  as}F_{n,l}^{\text{L}}-\omega_{ll^{\prime}%
}^{\left(  4\right)  as}F_{n,l^{\prime}}^{\text{L}}),\label{SBE-SSL}%
\end{gather}
which are of order of $V^{-2n+2}$, $V^{-2n+3}$ and $V^{-2n+4}$, respectively.
In linear response, $n=1$, these three equations just reduce to the familiar
ones in the study of the linear anomalous Hall effect \cite{Sinitsyn2008}. In
line with the Boltzmann recipe for the linear response
\cite{Sinitsyn2008,Xiao2017SOT-SBE}, the anti-symmetric part ($\omega
_{l^{\prime}l}^{as}\equiv\left(  \omega_{l^{\prime}l}-\omega_{ll^{\prime}%
}\right)  /2$) of $\omega_{ll^{\prime}}$, namely $\omega_{ll^{\prime}%
}^{\left(  3\right)  as}$ and $\omega_{ll^{\prime}}^{\left(  4\right)  as}$,
yields the skew scattering contribution to nonequilibrium phenomena, while the
inessential symmetric part of $\omega_{ll^{\prime}}^{\left(  3\right)  }$ and
$\omega_{ll^{\prime}}^{\left(  4\right)  }$ has been suppressed in the above
three equations.

$F_{n,l}^{\text{SL}}$ arises from the conventional skew scattering induced by
non-Gaussian disorder, thus can also be labeled by $F_{n,l}^{\text{csk}}$.
$F_{n,l}^{\text{SSL}}$ comprises contributions from the skew scattering
induced by Gaussian disorder (through $\omega_{l^{\prime}l}^{\left(  4\right)
as}$) and the electric-field-corrected scattering rate, thus can be decomposed
into
\begin{equation}
F_{n,l}^{\text{SSL}}=F_{n,l}^{\text{Gsk}}+F_{n,l}^{\text{a}},
\end{equation}
with%
\begin{align}
-\frac{e}{\hbar}\bm E\cdot\partial_{\bm k}F_{n-1,l}^{\text{Gsk}} &
=\sum_{l^{\prime}}\omega_{l^{\prime}l}^{\left(  2\right)  }(F_{n,l}%
^{\text{Gsk}}-F_{n,l^{\prime}}^{\text{Gsk}})\nonumber\\
&  +\sum_{l^{\prime}}(\omega_{l^{\prime}l}^{\left(  4\right)  as}%
F_{n,l}^{\text{L}}-\omega_{ll^{\prime}}^{\left(  4\right)  as}F_{n,l^{\prime}%
}^{\text{L}})
\end{align}
and
\begin{gather}
-\frac{e}{\hbar}\bm E\cdot\partial_{\bm k}F_{n-1,l}^{\text{a}}-\sum
_{l^{\prime}}\delta^{\bm E}\omega_{l^{\prime}l}^{\left(  2\right)  }%
(F_{n-1,l}^{\text{L}}-F_{n-1,l^{\prime}}^{\text{L}})=\nonumber\\
\sum_{l^{\prime}}\omega_{l^{\prime}l}^{\left(  2\right)  }(F_{n,l}^{\text{a}%
}-F_{n,l^{\prime}}^{\text{a}}).
\end{gather}
Here $F_{n,l}^{\text{a}}$ can be further decomposed as $F_{n,l}^{\text{a}%
}=F_{n,l}^{\text{a1}}+F_{n,l}^{\text{a2}}$, in correspondence to Eq.
(\ref{1+2}). We note again that $F_{1,l}^{\text{a2}}=0$ in the linear response.

\subsection{Electric current in the semiclassical framework}

It has been well known that $\bm v_{l}$ is not equal to the usual group
velocity $\bm v_{l}^{0}$, but contains corrections from interband virtual
transitions induced by both the electric-field and scattering
\cite{Xiao2017SOT-SBE,Sinitsyn2007}: $\bm v_{l}=\bm v_{l}^{0}+\bm
v_{l}^{\text{bc}}+\bm v_{l}^{\text{sj}}$. Here $\bm v_{l}^{\text{bc}}=\frac
{e}{\hbar}(\partial_{\bm{k}}\times\mathcal{A}_{l})\times\bm E$ and $\bm
v_{l}^{\text{sj}}=\sum_{l^{\prime}}\omega_{l^{\prime}l}^{\left(  2\right)
}\delta\bm r_{l^{\prime}l}$ are the Berry-curvature anomalous velocity and
side-jump velocity, respectively \cite{Sinitsyn2006}.

Therefore, the $n$th-order electric current is given by%
\begin{align}
\bm j_{n} &  =e\sum_{l}F_{n,l}^{\text{L}}\bm v_{l}^{0}+e\sum_{l}%
F_{n,l}^{\text{csk}}\bm v_{l}^{0}\nonumber\\
&  +e\sum_{l}F_{n,l}^{\text{Gsk}}\bm v_{l}^{0}+e\sum_{l}F_{n,l}^{\text{a1}}\bm
v_{l}^{0}+e\sum_{l}F_{n,l}^{\text{a2}}\bm v_{l}^{0}\nonumber\\
&  +e\sum_{l}F_{n,l}^{\text{L}}\bm v_{l}^{\text{sj}}+e\sum_{l}F_{n-1,l}%
^{\text{L}}\bm v_{l}^{\text{bc}}\label{semiclassical response}%
\end{align}
up to the first three leading orders of the weak disorder potential. The third
term on the second line is absent in all recent works on the semiclassical
Boltzmann theory of the second-order nonlinear anomalous Hall effect
\cite{Deyo2009,Konig2019,Fu2018,Du2018}.

Both $F_{n,l}^{\text{a1}}$ and $\bm v_{l}^{\text{sj}}$ are related to the
coordinate-shift, thereby the sum of these two terms is usually referred to as
the side-jump contribution
\cite{Deyo2009,Konig2019,Fu2018,Du2018,Sinitsyn2006,Sinitsyn2007,Sinitsyn2008}%
. However, $F_{n,l}^{\text{a1}}$ has nothing to do with the sideways shift,
which is the original meaning of side-jump \cite{Berger1970}. Accordingly, in
the following the terminology \textquotedblleft side-jump\textquotedblright%
\ is only assigned to the $\bm v_{l}^{\text{sj}}$ term.

\section{Model calculation in second-order nonlinear Hall effect}

To be specific, we illustrate the contribution from the $F_{n,l}^{\text{a2}}$
term in the second-order nonlinear Hall effect in inversion-breaking
nonmagnetic materials \cite{Fu2015,Ma2019,Mak2019}. To obtain analytic result,
we follow the previous publications involving $\delta_{1}^{\bm E}%
\omega_{l^{\prime}l}^{\left(  2\right)  }$ \cite{Konig2019,Du2018,Deyo2009}%
\ to take the constant relaxation time so that $\sum_{l^{\prime}}%
\omega_{l^{\prime}l}^{\left(  2\right)  }(F_{2,l}^{\text{a2}}-F_{2,l^{\prime}%
}^{\text{a2}})=F_{2,l}^{\text{a2}}/\tau$ and
\begin{equation}
F_{2,l}^{\text{a2}}=-\tau\sum_{\bm k^{\prime}}\frac{2\pi}{\hbar}\delta^{\bm
E}W_{l^{\prime}l}\delta(\epsilon_{l}-\epsilon_{l^{\prime}})(F_{1,l}^{\text{L}%
}-F_{1,l^{\prime}}^{\text{L}}).
\end{equation}
Here $F_{1,l}^{\text{L}}$ solves the conventional Boltzmann equation
(\ref{SBE-L}), reading $F_{1,l}^{\text{L}}=-\frac{e}{\hbar}\bm E\cdot
\partial_{\bm k}F_{0,l}^{\text{L}}\tau$ in the constant relaxation time
approximation. When the $\bm E$-field is applied in the $x$\ direction, the
resultant transverse current is
\begin{equation}
j_{y}^{a}=e\sum_{l}F_{2,l}^{\text{a2}}v_{l,y}^{0}\equiv\Xi_{yxx}^{a}E_{x}%
E_{x},
\end{equation}
where $\Xi_{yxx}^{a}$ is the corresponding second-order response coefficient.
One can show that $\Xi_{yxx}^{a}$ can be nonzero when the inversion symmetry
is broken, even if the time-reversal symmetry remains. This character is the
same as the known contributions of order of $\tau$ to the second-order
nonlinear Hall effect \cite{Konig2019,Fu2018,Du2018,Sodemann2019}%
.\begin{figure}[tbh]
\includegraphics[width=0.9\columnwidth]{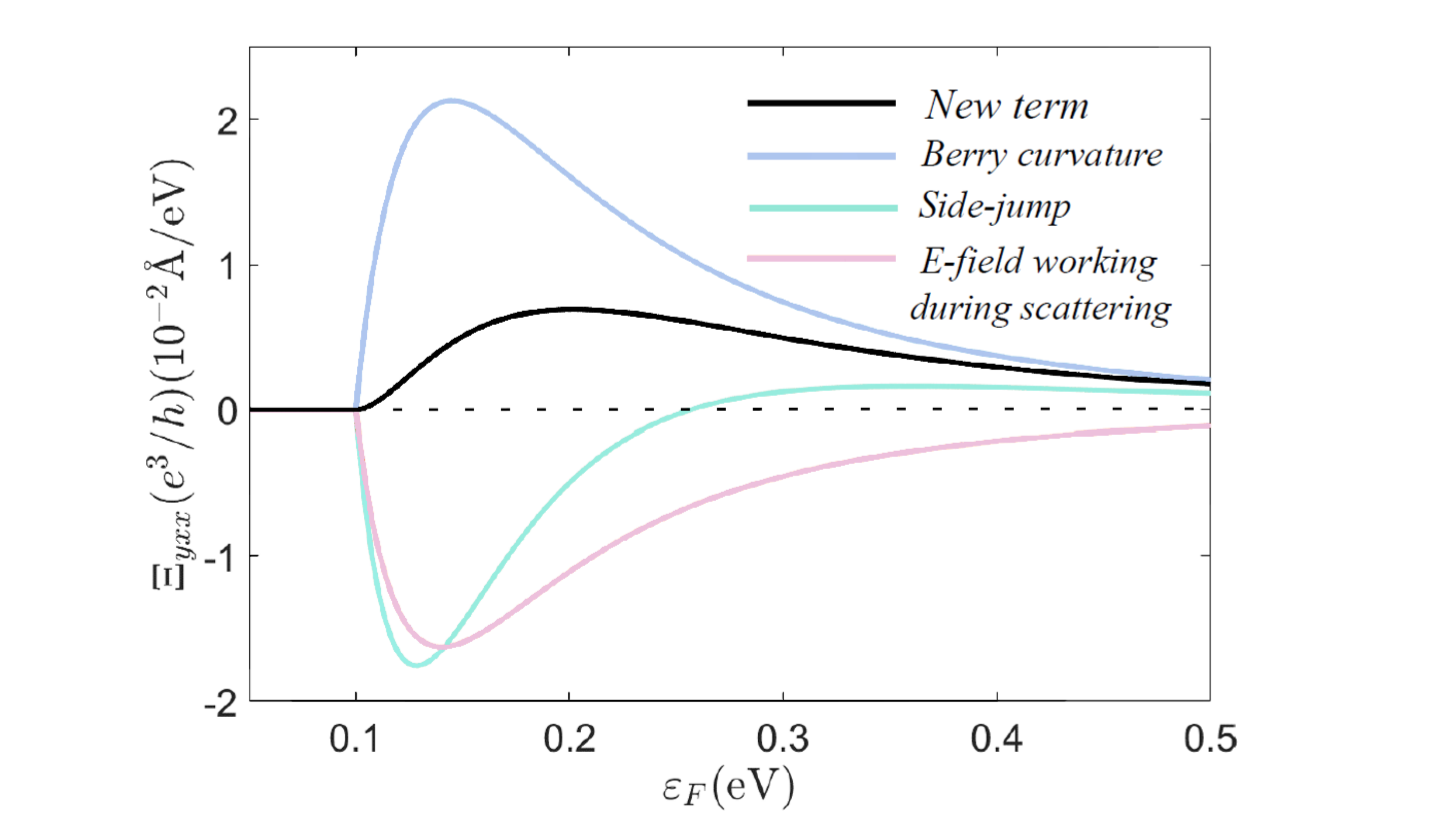}\caption{{}The
second-order nonlinear Hall responses in the 2D tilted massive Dirac model
that are beyond the conventional Boltzmann equation, from the Berry curvature
and side-jump velocities, $\bm E$-field working during scattering $e\bm
E\cdot\delta\bm r_{l^{\prime}l}$ and interband effect of $\bm E$-field during
scattering (the new term). Parameters are chosen as $t=0.1$ eV$\cdot$\r{A},
$v=1$ eV$\cdot$\r{A}, $\Delta=0.1$ eV, $n_{i}V_{0}^{2}=10^{2}$ (eV$\cdot$\r
{A})$^{2}$. $h=2\pi\hbar$ is the Planck constant.}%
\label{fig:Dirac-model}%
\end{figure}

Let us take the 2D tilted massive Dirac model \cite{Fu2015}
\begin{equation}
\hat{H}_{0}=tk_{x}+v\left(  k_{x}\sigma_{x}+k_{y}\sigma_{y}\right)
+\Delta\sigma_{z}%
\end{equation}
with scalar disorder as a concrete example, which is the minimal model of the
considered effect \cite{Fu2015,Du2018,Ma2019,Konig2019,Sodemann2019}.
$\sigma_{x,y,z}$ are the Pauli matrices, and the gapped Dirac cone is tilted
along the $x$ direction. Here one can consider the contribution from only one
Dirac cone because, as addressed in Refs. \cite{Fu2015,Konig2019}, taking into
account that from another one of the pair of Dirac cones simply doubles the
obtained result.

When the Fermi level only intersects the upper ($+$) band we have
\begin{equation}
\delta^{\bm E}W_{l^{\prime}l}=\frac{W_{\bm k^{\prime}\bm k}}{2}e\bm
E\cdot\left(  \frac{\Omega_{+\bm k^{\prime}}}{\Delta_{\bm k}}-\frac
{\Omega_{+\bm k}}{\Delta_{\bm k^{\prime}}}\right)  \hat{\bm{z}}\times\left(
\bm k^{\prime}-\bm k\right)  ,
\end{equation}
where $l=+\bm k$, $l^{\prime}=+\bm k^{\prime}$, $W_{\bm k^{\prime}\bm
k}=W_{l^{\prime}l}/|\langle u_{l}|u_{l^{\prime}}\rangle|^{2}$, $\Omega_{+\bm
k}$ is the Berry curvature of the upper band, and $\Delta_{\bm k}%
=\epsilon_{\bm k}^{+}-\epsilon_{\bm k}^{-}$. To compare with other
contributions obtained analytically for this model, we also assume weak
anisotropy $t\ll v$ and take \cite{Du2018,Sodemann2019} $1/\tau=n_{i}V_{0}%
^{2}\left(  \epsilon_{F}^{2}+3\Delta^{2}\right)  /\left(  4\hbar v^{2}%
\epsilon_{F}\right)  $ in the presence of pointlike impurities of density
$n_{i}$, for which $W_{\bm k^{\prime}\bm k}=n_{i}V_{0}^{2}$. It follows that
\begin{equation}
\Xi_{yxx}^{a}=-\frac{e^{3}}{2\pi\hbar}\frac{t\Delta}{n_{i}V_{0}^{2}}%
\frac{3v^{2}\left(  \epsilon_{F}^{2}-\Delta^{2}\right)  ^{2}}{\epsilon_{F}%
^{3}\left(  \epsilon_{F}^{2}+3\Delta^{2}\right)  ^{2}}%
\end{equation}
up to the first order of $t$. As shown in Fig. \ref{fig:Dirac-model},
$\Xi_{yxx}^{a}$ is of the similar magnitude to the previously identified
contributions \cite{Fu2015,Deyo2009,Du2018} that are also beyond the
conventional Boltzmann recipe \cite{note-dc}.

\section{Boltzmann transport emerging from quantum kinetics}

In this section we place the intuitive Boltzmann framework on the foundation
of quantum kinetics, by extending the density-matrix equation of motion
approach of Kohn and Luttinger \cite{KL1957,Luttinger1958} to nonlinear
responses.

\subsection{Basic formulations}

In the single-electron Hamiltonian $\hat{H}_{T}=\hat{H}_{0}+\hat{V}+\hat
{H}_{\bm E}$, $\hat{H}_{0}$ is the equilibrium disorder-free one, $\hat{V}$ is
the potential produced by randomly distributed impurities, and the $\bm
E$-field term $\hat{H}_{\bm E}=-e\bm E\cdot\bm re^{st}$ is switched on
adiabatically from the remote past $t=-\infty$. The physical situation is
obtained by taking the limit $s\rightarrow0^{+}$ \cite{KL1957}. In the case of
a weak $\bm E$-field, the single-particle density matrix is decomposed into
$\hat{\rho}_{T}=\sum_{n\geq0}\hat{\rho}_{n}$, where $\hat{\rho}_{0}$ is its
equilibrium value, $\hat{\rho}_{n}\propto E^{n}$ satisfies $\hat{\rho}%
_{n}\left(  t\rightarrow-\infty\right)  =0$ for $n\geq1$. Then the quantum
Liouville equation reduces to $[\hat{H}_{0}+\hat{V},\hat{\rho}_{0}]=0$ and
\begin{equation}
i\hbar\frac{\partial\hat{\rho}_{n}}{\partial t}=[\hat{H}_{\bm E},\hat{\rho
}_{n-1}]+[\hat{H}_{0}+\hat{V},\hat{\rho}_{n}],\text{ \ }\left(  n\geq1\right)
,
\end{equation}
where $[\hat{H}_{\bm E},\hat{\rho}_{n-1}]$ enables the recursion from
$\hat{\rho}_{n-1}$ to $\hat{\rho}_{n}$. Utilizing the ansatz \cite{KL1957}
$\hat{\rho}_{n}=\hat{f}_{n}e^{nst}$, where $\hat{f}=\sum_{n\geq0}\hat{f}_{n}$
is the single-particle density matrix at the time of interest $t=0$, in the
Bloch representation of $\hat{H}_{0}$ we have\ $\left(  n\geq1\right)  $
\begin{align}
\left(  \epsilon_{l}-\epsilon_{l^{\prime}}-i\hbar ns\right)  f_{n,ll^{\prime
}}  & =\sum_{l^{\prime\prime}}\left(  f_{n,ll^{\prime\prime}}V_{l^{\prime
\prime}l^{\prime}}-V_{ll^{\prime\prime}}f_{n,l^{\prime\prime}l^{\prime}%
}\right)  \nonumber\\
& +e\bm E\cdot\lbrack\bm r,\hat{f}_{n-1}]_{ll^{\prime}}.\label{EOM}%
\end{align}
When $l=l^{\prime}$ the equation of motion (\ref{EOM}) reduces to (more details in
Refs. \cite{KL1957,Xiao2018KL,Xiao2018JPCM})%
\begin{equation}
0=C_{n,l}+\sum_{l^{\prime}}^{\prime}\left(  f_{n,ll^{\prime}}V_{l^{\prime}%
l}-V_{ll^{\prime}}f_{n,l^{\prime}l}\right)  ,\label{EOM-diagonal}%
\end{equation}
otherwise we have ($f_{n,l}\equiv f_{n,ll}$)
\begin{align}
f_{n,ll^{\prime}} &  =\frac{C_{n,ll^{\prime}}}{\epsilon_{l}-\epsilon
_{l^{\prime}}-i\hbar ns}+\frac{f_{n,l}-f_{n,l^{\prime}}}{\epsilon_{l}%
-\epsilon_{l^{\prime}}-i\hbar ns}V_{ll^{\prime}}\nonumber\\
&  +\sum_{l^{\prime\prime}}^{\prime}\frac{f_{n,ll^{\prime\prime}}%
V_{l^{\prime\prime}l^{\prime}}-V_{ll^{\prime\prime}}f_{n,l^{\prime\prime
}l^{\prime}}}{\epsilon_{l}-\epsilon_{l^{\prime}}-i\hbar ns},\text{ \ \ (}l\neq
l^{\prime}\text{),}\label{EOM-off-diagonal}%
\end{align}
where $V_{ll}$ is absorbed into $H_{0}$ and then $V_{ll}=0$ \cite{KL1957}.
Here
\begin{equation}
C_{n,l}=e\bm E\cdot\{i\partial_{\bm{k}}f_{n-1,l}+[\mathcal{A},\hat{f}%
_{n-1}]_{ll}\},
\end{equation}
and%
\begin{equation}
C_{n,ll^{\prime}}=e\bm E\cdot\{i\left(  \partial_{\bm{k}}+\partial
_{\bm{k}^{\prime}}\right)  f_{n-1,ll^{\prime}}+[\mathcal{A},\hat{f}%
_{n-1}]_{ll^{\prime}}\}.
\end{equation}

According to Eq. (\ref{EOM-off-diagonal}), in the case of weak disorder
potential $f_{n,ll^{\prime}}$\ is generally one order of $V$ higher than
$f_{n,l}$. Thereby, Eq. (\ref{EOM-off-diagonal}) can be solved by an iterative
procedure, which yields the expression for $f_{n,ll^{^{\prime}}}$\ in terms of
$f_{n,l}$ \cite{KL1957,Luttinger1958}. Substituting this solution into Eq.
(\ref{EOM-diagonal}) leads to an equation only concerning the diagonal element
$f_{n,l}$ for $n\geq1$. The disorder average of this latter equation yields a
Boltzmann-type equation for $\langle f_{n,l}\rangle_{c}$, provided that one
assumes $f_{n,l}$ is self-averaged, i.e., $\langle f_{n,l}VV\rangle
_{c}=\langle f_{n,l}\rangle_{c}\langle VV\rangle_{c}$. This assumption plays
the similar role to that of the assumption of molecular chaos in deriving the
classical Boltzmann equation from the classical Liouville equation
\cite{Kardar}. One can then identify $\langle f_{n,l}\rangle_{c}$ with the
occupation function used in the Boltzmann framework
\begin{equation}
F_{n,l}\equiv\langle f_{n,l}\rangle_{c},\text{ \ }F_{l}\equiv\langle
f_{l}\rangle_{c}=\sum_{n\geq0}\langle f_{n,l}\rangle_{c}.
\end{equation}

On the other hand, the $n$th-order charge current in the density-matrix
formulation is given by%
\begin{equation}
\bm j_{n}=\bm j_{n}^{\text{d}}+\bm j_{n}^{\text{od}},\label{quantum response}%
\end{equation}
where $\bm j_{n}^{\text{d}}=e\sum_{l}\left\langle f_{n,l}\right\rangle _{c}\bm
v_{l}^{0}$ and $\bm j_{n}^{\text{od}}=e\sum_{ll^{^{\prime}}}^{^{\prime}%
}\langle f_{n,ll^{^{\prime}}}\rangle_{c}\bm v_{l^{^{\prime}}l}^{0}$\ are the
band-diagonal and band-off-diagonal (the matrix element of velocity operator
$\bm v_{l^{\prime}l}^{0}$ is diagonal in $\bm k$) responses, respectively.

\subsection{Band-diagonal response  }

We first illustrate the aforementioned iterative procedure in its lowest
order, where $C_{n,l}=ie\bm E\cdot\partial_{\bm{k}}f_{n-1,l}$, and
$f_{n,ll^{\prime}}$ is given by the second term on the right-hand-side of Eq.
(\ref{EOM-off-diagonal}). Plugging them into Eq. (\ref{EOM-diagonal}) leads
to, after disorder average, the most conventional Boltzmann equation
(\ref{SBE-L}). Then, up to the third-order iteration the Boltzmann equations
(\ref{SBE-SL}) and (\ref{SBE-SSL}) also emerge after the disorder average,
thus%
\begin{align}
\bm j_{n}^{\text{d}} &  =e\sum_{l}F_{n,l}^{\text{L}}\bm v_{l}^{0}+e\sum
_{l}F_{n,l}^{\text{csk}}\bm v_{l}^{0}\nonumber\\
&  +e\sum_{l}F_{n,l}^{\text{Gsk}}\bm v_{l}^{0}+e\sum_{l}F_{n,l}^{\text{a1}}\bm
v_{l}^{0}+e\sum_{l}F_{n,l}^{\text{a2}}\bm v_{l}^{0}.\label{d}%
\end{align}
Most details of the iteration procedure have in fact already been presented in
previous papers \cite{KL1957,Xiao2018KL,Luttinger1958,Xiao2018JPCM}, and are
also provided in the Supplemental Material \cite{Supp} for the convenience of
the interested readers.

It is apparent that in the higher orders of the iteration $C_{n,l}$ contains
the combination effect of the electric field and disorder, which leads finally
to the additional driving term related to $\delta^{\bm E}\omega_{l^{\prime}%
l}^{\left(  2\right)  }$. In the linear anomalous Hall effect, only the
coordinate-shift-related component, namely $\delta_{1}^{\bm E}\omega
_{l^{\prime}l}^{\left(  2\right)  }$, survives in the resulting Boltzmann
equation, as has been elaborated in Refs.
\cite{Xiao2018KL,Luttinger1958,Xiao2018JPCM}. On the other hand, in nonlinear
responses the additional driving term related to $\delta_{2}^{\bm E}%
\omega_{l^{\prime}l}^{\left(  2\right)  }$ is derived from the quantum
kinetics for the first time, as is detailed in the Supplemental Material
\cite{Supp}.

Lastly, we emphasize that the equilibrium density matrix deserves separate
discussions. It should be obtained from the definition of the single-particle
density matrix \cite{Supp}, and the leading value is $f_{0,ll^{\prime}}%
=\delta_{ll^{\prime}}f_{0,l}$, with $f_{0,l}$\ the Fermi distribution. Note
that the equilibrium density matrix is also altered by disorder and thus does
not coincide with $f_{0,l}$. Through the $C_{1,ll^{\prime}}$ term, the
disorder-induced corrections to $f_{0,ll^{^{\prime}}}$ incorporate the effect
of the $\bm E$-field during scattering into linear response. The neglect of
this fact would lead to the absence of the $e\bm E\cdot\delta\bm r_{l^{\prime
}l}$ contribution to $f_{1,l}$. In fact this is one of the main differences
between the Kohn-Luttinger approach and another quantum kinetic approach
employed recently to study the linear and nonlinear anomalous Hall effects
\cite{Sodemann2019,Culcer2017}. More detailed discussions on this issue are
presented later.

\subsection{Band-off-diagonal response}

The leading nonzero contribution to the off-diagonal response $\bm
j_{n}^{\text{od}}$ is of $O(V^{-2n+2})$, given by the second-order iteration
of Eq. (\ref{EOM-off-diagonal}):%
\begin{gather}
\left\langle f_{n,ll^{\prime}}\right\rangle _{c}=\frac{e\bm E\cdot
\mathcal{A}_{ll^{\prime}}(F_{n-1,l^{\prime}}^{\text{L}}-F_{n-1,l}^{\text{L}}%
)}{\epsilon_{l}-\epsilon_{l^{\prime}}-i\hbar ns}\\
+\sum_{l^{\prime\prime}}^{\prime}\frac{\left\langle V_{ll^{\prime\prime}%
}V_{l^{\prime\prime}l^{\prime}}\right\rangle _{c}}{\epsilon_{l}-\epsilon
_{l^{\prime}}-i\hbar ns}[\frac{F_{n,l}^{\text{L}}-F_{n,l^{\prime\prime}%
}^{\text{L}}}{\epsilon_{l}-\epsilon_{l^{\prime\prime}}-i\hbar ns}%
-\frac{F_{n,l^{\prime\prime}}^{\text{L}}-F_{n,l^{\prime}}^{\text{L}}}%
{\epsilon_{l^{\prime\prime}}-\epsilon_{l^{\prime}}-i\hbar ns}],\nonumber
\end{gather}
and can be readily cast into \cite{Supp}%
\begin{equation}
\bm j_{n}^{\text{od}}=e\sum_{l}F_{n-1,l}^{\text{L}}\bm v_{l}^{\text{bc}}%
+e\sum_{l}F_{n,l}^{\text{L}}\bm v_{l}^{\text{sj}}.\label{od}%
\end{equation}
Here $\bm v_{l}^{\text{bc}}$ and $\bm v_{l}^{\text{sj}}$ coincide respectively
with the Berry-curvature anomalous velocity and side-jump velocity
\cite{Xiao2017SOT-SBE}.

Summing up, our quantum kinetic theory shows that arbitrary $n$th-order
response retains the same form of Eqs. (\ref{quantum response}), (\ref{d}) and
(\ref{od}), namely the semiclassical Boltzmann result Eq.
(\ref{semiclassical response}), up to the first three leading-order
contributions in the weak disorder potential. The correspondence of the basic
ingredients in the Boltzmann theory to the density matrix response is
summarized in Table \ref{correspondence}. It is worthwhile to remind here that
interband virtual processes play the essential role also in the band-diagonal
response of the density matrix.\begin{table}[t]
\caption{Correspondence of Boltzmann transport to the band-off-diagonal (od)
and band-diagonal (d) responses of density matrix. $\bm v_{l}^{\text{bc}}$ and
$\bm v_{l}^{\text{sj}}$ are the Berry-curvature and side-jump velocities,
respectively. $\delta^{\bm E}\omega_{l^{\prime}l}^{\left(  2\right)  }$ is the
$\bm E$-field corrected scattering rate. $\omega_{ll^{\prime}}^{\left(
3\right)  as}$ and $\omega_{ll^{\prime}}^{\left(  4\right)  as}$ yield the
skew scattering.}
\centering%
\begin{tabular}
[c]{|c|c|}\hline
semiclassical ingredients & density matrix response\\\hline
$\bm v_{l}^{\text{bc}}$ & $\bm j_{n}^{\text{od}}$\\\hline
$\bm v_{l}^{\text{sj}}$ & $\bm j_{n}^{\text{od}}$\\\hline
$\delta^{\bm E}\omega_{l^{\prime}l}^{\left(  2\right)  }=\delta_{1}^{\bm
E}\omega_{l^{\prime}l}^{\left(  2\right)  }+\delta_{2}^{\bm E}\omega
_{l^{\prime}l}^{\left(  2\right)  }$ & $\bm j_{n}^{\text{d}}$\\\hline
$\omega_{ll^{\prime}}^{\left(  3\right)  as},\omega_{ll^{\prime}}^{\left(
4\right)  as}$ & $\bm j_{n}^{\text{d}}$\\\hline
\end{tabular}
\label{correspondence}%
\end{table}

\section{Comparison to other theories}

We start by noting that, the similar idea to the intuitive consideration
leading to the $\bm E$-field-influenced scattering matrix element Eq.
(\ref{new term}) also appeared in two publications by Tarasenko
\cite{Tarasenko2011,Tarasenko2007}. In these two papers a nonlinear current
arises due to the $\bm E$-field-induced admixture of excited conduction- and
valence-band states to the ground-subband wave function in quantum wells. In
fact one can find that our Eqs. (\ref{mixture}), (\ref{W}) and (\ref{new term}%
) are quite similar to Eqs. (13) -- (15) in Ref. \cite{Tarasenko2011}. The
difference is that, in Refs. \cite{Tarasenko2011,Tarasenko2007} the
electric-field component $E_{z}$ (in the z direction of the quantum well)
mixes the quantum-confined states, whereas in the present work the electric
field $E_{x}$ mixes the Bloch states of electrons.

Next we compare our theory to the previous works on the Boltzmann formulation
of the nonlinear Hall effect \cite{Deyo2009,Konig2019,Fu2018,Du2018}. All
these works just generalized the Boltzmann theory for the linear anomalous
Hall effect \cite{Sinitsyn2008} directly and phenomenologically into the
nonlinear response. First, the $\delta_{2}^{\bm E}\omega_{l^{\prime}%
l}^{\left(  2\right)  }$ term proposed in the present study has no counterpart
in the linear response and thus is beyond such direct generalization of the
linear theory. Second, Our quantum theory supports the form of the
coordinate-shift-related $\delta_{1}^{\bm E}\omega_{l^{\prime}l}^{\left(
2\right)  }$ speculated intuitively in Refs. \cite{Deyo2009,Konig2019,Du2018},
which differs from the one proposed in Ref. \cite{Fu2018}.

Our theory is also different from the other quantum kinetic one of the
nonlinear Hall effect posted recently \cite{Sodemann2019}. This latter theory
is based on the nonlinear generalization of a linear-response density-matrix
theory \cite{Culcer2017}. Thus in the following we discuss first the
difference between this linear-response theory and ours, and then that between
the theory of Ref. \cite{Sodemann2019} and ours.

The whole second line of Eq. (\ref{d}), which arises from the band-diagonal
response of the density matrix, is missed in the linear-response theory of
Ref. \cite{Culcer2017}. Firstly, in this theory the equilibrium density matrix
is identified to be just the Fermi distribution. However, as we have stressed
in the last paragraph of Sec. IV. B, the equilibrium density matrix is not
equal to the Fermi distribution and has the disorder-induced correction. It is
this correction that leads finally to the coordinate-shift-related $\delta
_{1}^{\bm E}\omega_{l^{\prime}l}^{\left(  2\right)  }$ and thus to
$F_{1,l}^{\text{a1}}$ \cite{Xiao2018KL,Luttinger1958,Xiao2018JPCM}. Secondly,
the theory of Ref. \cite{Culcer2017} only considers the lowest-order Born
approximation in calculating the scattering rate, thus $F_{1,l}^{\text{Gsk}}$
is missed. The theory of Ref. \cite{Culcer2017} was shown to work well for the
linear anomalous Hall effect in the spin-polarized Rashba model and for the
spin Hall effect in the Rashba model. However, the peculiarity of the Rashba
models in fact plays the basic role in this success: in the spin-polarized
Rashba model $F_{n,l}^{\text{Gsk}}+F_{n,l}^{\text{a1}}=0$ in the case of
scalar point-like impurities within the noncrossing approximation for
$F_{n,l}^{\text{Gsk}}$ \cite{Xiao2017AHE}, whereas in the Rashba model the
diagonal element of the spin-current operator $\bm j^{s}$\ in the Bloch
representation is zero ($\left(  \bm j^{s}\right)  _{l}^{0}=0$)
\cite{Xiao2017SHE}. Therefore, when applied to another model, like the
two-dimensional gapped Dirac model, one can check that the theory of Ref.
\cite{Culcer2017} cannot reproduce the same anomalous Hall conductivity as the
previous theories \cite{Sinitsyn2007}.

Now we turn to the theory of Ref. \cite{Sodemann2019}. In the case of linear
response, this theory still misses the disorder induced correction to the
equilibrium density matrix, thus misses the contribution from $F_{1,l}%
^{\text{a1}}$. To be more specific, one can check that, the side-jump
conductivity in the first equation of Eq. (26) of Ref. \cite{Sodemann2019} is
in fact only one half of the side-jump conductivity defined in Ref.
\cite{Sinitsyn2007}. Another half, namely the contribution from $F_{1,l}%
^{\text{a1}}$, disappears: it is contained in neither the first nor the second
equation of Eq. (26) of Ref. \cite{Sodemann2019}.

Because the linear-response density matrix is vital in producing the
nonlinear-response one, the aforementioned difference makes the nonlinear
theory of Ref. \cite{Sodemann2019} also different from ours. While the
band-off-diagonal response Eq. (\ref{od}) is produced in Ref.
\cite{Sodemann2019}, the second line of the band-diagonal response Eq.
(\ref{d}) is not. This means that only the Berry-curvature dipole and
side-jump velocity contributions to the second-order nonlinear Hall effect
proposed in the previous semiclassical theory
\cite{Deyo2009,Konig2019,Fu2018,Du2018} have been identified in the quantum
kinetic theory of Ref. \cite{Sodemann2019}. At the present stage only our
theory establishes the consistency between the renewed Boltzmann and quantum
kinetics in nonlinear responses.

\section{Conclusion}

In conclusion, we have proposed a modified Boltzmann framework for nonlinear
electric-transport, and identified an interband-coherence effect induced by dc
electric fields during scattering. This effect has no counterpart in linear
response, and thus is missed in the previous nonlinear Boltzmann formalism for
the nonlinear Hall effect which is just the direct generalization of the
linear Boltzmann theory. The proposed Boltzmann formulation has been confirmed
by a quantum kinetic theory. This theory also shows that arbitrary $n$th-order
nonlinear response to a dc electric field, up to the first three leading
contributions in the weak disorder potential, is handled by the same few
gauge-invariant semiclassical ingredients.

\begin{acknowledgments}
We thank T. Chai for helpful discussions. Q.N. is supported by DOE (DE-FG03-02ER45958, Division of Materials Science and Engineering) on the transport formulation in this work. C.X. is supported by NSF (EFMA-1641101) and Welch Foundation (F-1255).
\end{acknowledgments}

\end{document}